\documentstyle[twoside,fleqn,espcrc2,epsfig]{article}

\newcommand{\AmS}{{\protect\the\textfont2
  A\kern-.1667em\lower.5ex\hbox{M}\kern-.125emS}}

\title{Determining gravitational lensing effects on supernovae observations}

\author{Edvard M\"ortsell\address{Department of Physics, 
    Stockholm University, \\ 
        Box 6730, S-113 85 Stockholm, Sweden}%
        \thanks{E-mail address: edvard@physto.se.}}

\begin{document}

\begin{abstract}
In this paper we use a generalized version of a method originally proposed 
by Holz and Wald \cite{art:HolzWald1998} to investigate the effects from 
gravitational lensing on Type Ia supernovae measurements. 
We find that results for different mass distributions in smooth dark matter halos
are very similar, making lensing effects predictable for a broad range of density 
profiles.
Also, a sample of 100 supernovae at $z\sim 1$, should be sufficient
to discriminate between the case of all dark matter in smooth halos
and the extreme case of all dark matter in point-like objects. 

\end{abstract}

\maketitle

\section{Introduction}
Gravitational lensing has become an increasingly important tool in 
astrophysics and cosmology. In particular, the effects of lensing has
to be taken into account when studying sources at high redshifts. 
In an inhomogeneous universe, sources may be magnified or 
demagnified with respect to the case of a homogeneous universe with the same
average energy density.

Holz and Wald (HW; \cite{art:HolzWald1998}) have presented a
method for determining gravitational lensing effects in
inhomogeneous universes. Their use of realistic galaxy models has
been limited to the singular, truncated isothermal sphere (SIS) with a
fixed mass. We have generalized their method to allow for
matter distributions more accurately describing the actual properties 
of galaxies. The list of matter distributions have been extended to
include the density profile proposed by Navarro, Frenk and 
White (NFW; \cite{art:NFW}) and we use a distribution of galaxy masses. 
Also, other matter distribution parameters such as the scale radius
of the NFW halo and the cut-off radius of the SIS halo are determined
from distributions reflecting real galaxy properties. For further details 
we refer to Bergstr\"om {\it et al.} \cite{MGEM-lens} where also
the method of HW has been generalized to allow for general perfect fluids
with non-vanishing pressure.

As an application of the method, we investigate gravitational lensing effects on
observations of distant supernovae. Specifically, we consider the
effects on supernova luminosity distributions.

\section{Method}
\begin{figure}[t]
  \centerline{\hbox{\epsfig{figure=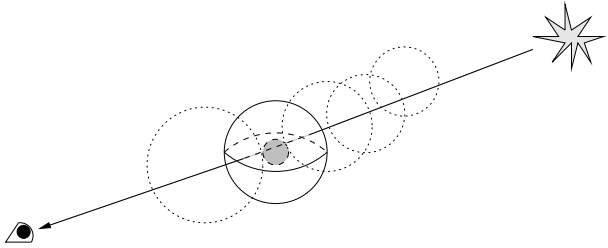,width=0.5\textwidth}}}
  \caption{In the method of HW, light rays are traced backwards to the
desired redshift by being sent through a series of cells, each time
with a randomly selected impact parameter.}\label{fig:cell2} 
\end{figure}

The method of HW can be summarized as follows: 
First, a Friedmann-Lema\^{\i}tre (FL) background geometry is
selected. Inhomogeneities are accounted for by specifying matter
distributions in cells with energy density equal to that of the
underlying FL model. A light ray is traced backwards to the
desired redshift by being sent through a series of cells, each time
with a randomly selected impact parameter. After each cell, the
FL background is used to update the scale factor
and expansion (see Fig~\ref{fig:cell2}). 
By using Monte Carlo techniques to trace a large number
of light rays, and by appropriate weighting \cite{art:HolzWald1998,MGEM-lens},
statistics for the apparent luminosity of the source is obtained.

\section{Gravitational lensing of supernovae}
One of the major goals of cosmology is to determine the cosmological
parameters of our Universe. It has been realised  
that observations of supernovae at high redshift can be
used for this purpose, in particular for determining the value of the cosmological
constant. In fact, several collaborations with this in mind are in
progress, and the first sets of data show an intriguing hint of a
non-vanishing cosmological constant \cite{art:Perlmutter-et-al1999,art:Riess-et-al1998}. 

\begin{figure}[t]
  \centerline{\hbox{\epsfig{figure=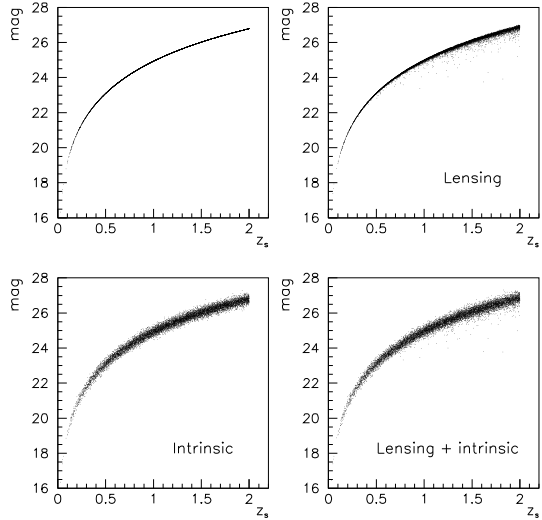,width=0.5\textwidth}}}
  \caption{Luminosity distribution for 10\,000 sources in an 
    $\Omega_M=0.3$, $\Omega_\Lambda=0.7$ universe with 80 \% isothermal spheres and 
    20 \% compact objects. Note that lensing effects become comparable to the
    intrinsic dispersion of Type Ia supernovae at redshifts larger than one.}
  \label{fig:dispersion} 
\end{figure}

Although the two groups which have published results mutually agree 
on the best-fit parameters, it is important to note that the effects
of geometry are small (on the order of half a magnitude), and the need
to go to even higher redshift to get larger effects is obvious.
When observing such distant sources, at redshift greater than 
unity, it is necessary to estimate the effects of lensing due to
inhomogeneities in the matter distribution. In Fig.~\ref{fig:dispersion}, 
we compare the effects from gravitational lensing with the intrinsic dispersion 
of Type Ia supernovae. It is evident that gravitational effects become comparable 
to the intrinsic dispersion at redshifts larger than one. 

Of course, the additional dispersion caused by graviational lensing will be a 
source of systematic error in the cosmological parameter determination with 
Type Ia supernovae. However, a possible virtue of lensing is that the 
distribution of luminosities might be used to obtain some information on the 
matter distribution in the Universe, see Goliath and M\"ortsell \cite{MGEM-letter}. 

\section{Results}\label{sec:results}

\begin{figure}[t]
  \centerline{\hbox{\epsfig{figure=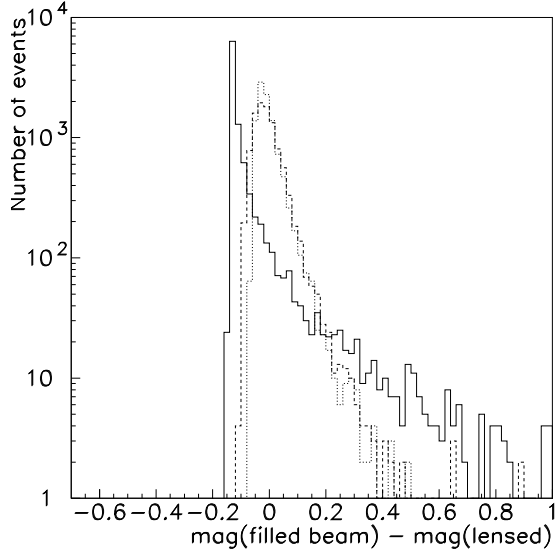,width=0.5\textwidth}}}
  \caption{Luminosity distributions for 10\,000 perfect
    standard candles at
    redshift $z=1$ in an $\Omega_M=0.3$, $\Omega_\Lambda=0.7$
    universe. The magnification zero point is the luminosity in the
    corresponding homogeneous (``filled-beam'') model. 
    The full line corresponds to the point-mass case; the
    dashed line is the distribution for SIS halos, and the dotted line
    is the NFW case.}\label{fig:model-2}
\end{figure}

In Fig.~\ref{fig:model-2}, we compare the luminosity  
distributions obtained with point masses, SIS lenses and NFW matter 
distributions in an $\Omega_M=0.3$, $\Omega_\Lambda=0.7$ universe,
currently favoured by Type Ia supernova measurements 
\cite{art:Perlmutter-et-al1999,art:Riess-et-al1998} and anisotropy 
measurements of the cosmological background radiation \cite{boomerang,maxima}.
Sources are assumed to be perfect standard candles.
The magnification, given in magnitudes, has its zero point at the filled beam 
value, i.e., the value one would get in a homogeneous universe. 
Note that negative values corresponds
to demagnifications and positive values to magnifications. 
Results for SIS halos and NFW halos are very similar, even when we have
no intrinsic luminosity dispersion of the sources.

\begin{figure}[t]
  \centerline{\hbox{\epsfig{figure=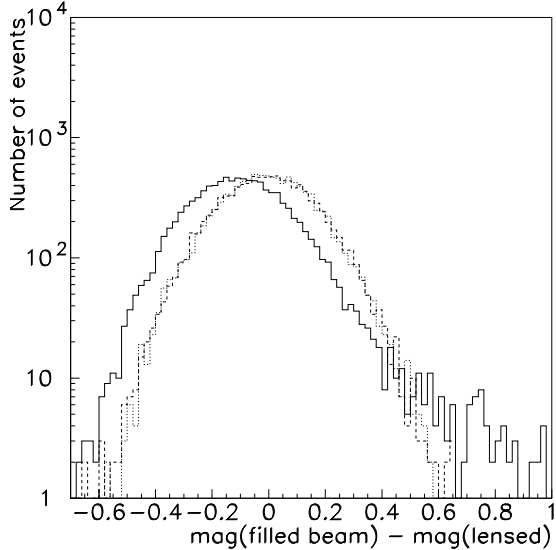,width=0.5\textwidth}}}
  \caption{Luminosity distributions for 10\,000 sources at
    redshift $z=1$ in an $\Omega_M=0.3$, $\Omega_\Lambda=0.7$
    universe. This is the same situation as depicted in
    Fig.~\ref{fig:model-2}, only that we have added an intrinsic 
    luminosity dispersion of the sources with 
    $\sigma_m = 0.16$ mag. (corresponding to the case of Type Ia
    supernovae).}\label{fig:intsig-2} 
\end{figure}

In Fig.~\ref{fig:intsig-2} we have added an intrinsic luminosity
dispersion represented by a Gaussian distribution with $\sigma_m = 0.16$
mag., due to the fact that Type Ia supernovae are not perfect standard candles. 
The effect is to make the characteristics of the luminosity
distributions even less pronounced, since the form of the resulting
luminosity distributions predominantly is determined by the form of
the intrinsic luminosity distribution. It is still possible to
observationally distinguish whether lenses consist of compact
objects or smooth galaxy halos (see also \cite{MetcalfSilk,SeljakHolz}).
Generating several samples containing 100 
supernova events at $z=1$ in an $\Omega_M=0.3$, $\Omega_\Lambda=0.7$
cosmology filled with smooth galaxy halos, we find that for 98 \% of the
samples one can rule out a point-mass distribution with  99 \%
confidence level\footnote{However, in 1 \% of the samples, we will
  erroneously rule out the halo distribution with the same confidence
  level.}. Furthermore, for a similar sample containing 200
supernovae, the confidence level is increased to 99.99 \%.

A more extensive discussion of the luminosity distributions of perfect
standard candles obtained 
with the different halo models at different source redshifts can be 
found in \cite{MGEM-lens}, where also some analytical
fitting formulas for the probability distributions are given.

\section{Summary}

We have generalized the method of Holz and Wald \cite{art:HolzWald1998} to allow
for matter distributions reflecting the actual properties of galaxies. 
One of the virtues of this method is that it can be continuously refined
as one gains more information about the matter distribution in the 
universe from observations. The motivation for these generalizations
is to use this method as part of a model for simulation of
high-redshift supernova observations, 
{\em the SuperNova Observation Calculator (SNOC)} .

In this paper, we have considered lensing effects on supernova luminosity 
distributions. 
Results for different mass distributions in smooth dark matter halos
were found to be very similar, making lensing effects predictable for
a broad range of density profiles.
Furthermore, given a sample of 100 supernovae at $z\sim 1$, one should
be able to discriminate between the case with smooth dark matter halos
and the (unlikely) case of 
having a dominant component of dark matter in point-like objects.

\end{document}